\documentstyle[12pt]{article}
\input epsf.sty
\def\ZZZ{{\hbox{ Z\kern-1.6mm Z}}}

\newcommand{\MM}{{\cal M}}

\newcommand{\LL}{{\cal L}}

\newcommand{\wt}{\widetilde}
\newcommand{\wh}{\widehat}

\newcommand{\NN}{{\cal N}}

\newcommand{\SSS}{{\cal S}}

\newcommand{\be}{\begin{equation}}
\newcommand{\ee}{\end{equation}}
\newcommand{\ben}{\begin{eqnarray}\displaystyle}
\newcommand{\een}{\end{eqnarray}}
\newcommand{\refb}[1]{(\ref{#1})}
\newcommand{\p}{\partial}
\newcommand{\sectiono}[1]{\section{#1}\setcounter{equation}{0}}

\def\one{{\hbox{ 1\kern-.8mm l}}}
\def\zero{{\hbox{ 0\kern-1.5mm 0}}}

\begin{document}
{}~
{}~
\hfill\vbox{\hbox{hep-th/0508042}
}\break

\vskip .6cm

{\baselineskip20pt
\begin{center}
{\Large \bf
Entropy Function for Heterotic Black Holes
} 

\end{center} }

\vskip .6cm
\medskip

\vspace*{4.0ex}

\centerline{\large \rm
Ashoke Sen}

\vspace*{4.0ex}

\centerline{\large \it Harish-Chandra Research Institute}

\centerline{\large \it  Chhatnag Road, Jhusi,
Allahabad 211019, INDIA}

\vspace*{1.0ex}

\centerline{E-mail: ashoke.sen@cern.ch,
sen@mri.ernet.in}

\vspace*{5.0ex}

\centerline{\bf Abstract} \bigskip

We use the entropy function formalism to study the effect of the
Gauss-Bonnet term on the entropy of spherically symmetric extremal black
holes in heterotic string theory in four dimensions. Surprisingly the
resulting entropy and the near horizon metric, gauge field strengths
and the axion-dilaton field
are identical to those obtained by Cardoso et. al. for
a supersymmetric version of the
theory that contains Weyl tensor squared term instead of the Gauss-Bonnet
term. We also study the effect of holomorphic anomaly on the entropy using
our formalism. Again the resulting attractor equations for the axion-dilaton
field
and the black hole
entropy agree with the corresponding equations for the supersymmetric version
of the theory. These
results suggest that there
might be a simpler description of supergravity with curvature squared
terms in which we supersymmetrize the Gauss-Bonnet term instead of the
Weyl tensor squared term.
 
\vfill \eject

\baselineskip=18pt

\tableofcontents

\sectiono{Introduction and Summary} \label{s1}

In a previous paper\cite{0506177} we developed a simple method for 
computing the entropy of a spherically symmetric extremal black hole in a 
theory of gravity coupled to abelian gauge fields, 
neutral scalar fields (and possibly other anti-symmetric tensor fields in
dimension 
$>4$) with arbitary higher derivative interactions.
In particular we gave an algorithm for constructing an `entropy 
function', -- a
function of the parameters labelling the near horizon background, -- such 
that 
extremization of this function with respect to the parameters determines 
the correct values of these parameters for a given set of charges carried 
by the black hole. Furthermore the value of this function at the 
extremum gives the entropy of the black hole. Related (but complementary) 
results have been obtained in \cite{0506176,0507096}.

In this paper we apply this method to study the effect of higher 
derivative terms on the entropy of various 
extremal black 
holes in four dimensional heterotic string theory. More specifically we 
add to 
the usual 
supergravity action a Gauss-Bonnet term that is known to arise in tree 
level heterotic string theory\cite{rzwiebach,9610237}, 
and analyze the black hole entropy in the 
resulting theory. This problem was studied earlier in \cite{9711053} in an 
approximation where the modification of the near horizon geometry due to 
the Gauss-Bonnet term was ignored, and only 
the additional contribution of the Gauss-Bonnet term 
to the expression for the entropy was taken into account. A somewhat
different  scheme, where again we do not explicitly take
into account the effect of backreaction of the Gauss-bonnet term on the
near horizon geometry, has been suggested in \cite{0506251}.
The entropy function 
formalism allows us to go beyond these approximations.

During this study we find some surprises. 
Refs.\cite{9801081,9812082,9904005,9906094,9910179,0007195,0009234,0012232} 
studied a closely related 
theory, where instead of the Gauss-Bonnet term we add to the action a term 
proportional to the square of the Weyl tensor and 
infinite number of other terms required for supersymmetric completion of 
the action. The coefficient function of the Weyl tensor squared term is 
adjusted so that the term involving the square of the
Riemann tensor has the same coefficient in both theories. Black hole 
entropy in these supersymmetric theories was computed using a 
completely different method. 
In the appropriate approximation the results of these computations turned
out to agree with those of \cite{9711053,0506251}.
We find that after taking into account the effect of backreaction our results 
for not only the 
black hole entropy, but also the near horizon metric, gauge field 
strengths and the axion-dilaton field, agree with those of 
\cite{9801081,9812082,9904005,9906094,9910179,0007195,0009234,0012232}. 
This exact agreement between the two sets of results is 
surprising considering that we do not even have a fully supersymmetric 
action. 
This perhaps suggests that there is a simpler way to 
supersymmetrize curvature squared terms in the action based on the 
Gauss-Bonnet combination rather than the Weyl tensor squared term. It will 
be interesting to explore this possibility.

Heterotic string theory on $T^6$ and more general CHL compactifications 
discussed in \cite{9505054,9506048,9507027,9507050,9508144,9508154} have 
S-duality symmetry group SL(2,\ZZZ ) (as in the case of toroidal 
compactification\cite{9402002}) or a subgroup of SL(2,\ZZZ )
(as in the case of CHL models\cite{9507050}). The tree level 
curvature squared terms are not invariant under this S-duality group, and
additional terms are needed to restore the S-duality invariance of the
action.
This 
amounts to changing the coefficient of the curvature squared term to an 
S-duality invariant function of the axion-dilaton field. If this  
function can be regarded as the imaginary part of a
holomorphic function, then the action based on the Weyl tensor squared 
term can be supersymmetrized and the resulting correction to the black 
hole entropy can be computed\cite{9801081,9812082,9904005,
9906094,9910179,0007195,
0009234,0012232}. Generically however this holomorphicity 
requirement is not compatible with the requirement of S-duality and the 
coefficient of the curvature squared term contains a part that cannot be 
regarded as the imaginary part of a holomorphic 
function\cite{9302103,9307158,9309140}. In such cases it is not known how 
to supersymmetrize this term, and hence the method of 
\cite{9801081,9812082,9904005,9906094,9910179,0007195,0009234,0012232} is 
not directly applicable. Nevertheless a form of the modified entropy and 
attractor equations was guessed in \cite{9906094,0007195} by requiring
S-duality invariance of the final result.
On the other hand since in our analysis we do not supersymmetrize 
the action and just work with the Gauss-Bonnet term, 
we do not have any 
difficulty in extending our analysis to include the non-holomorphic terms, 
and find an expression for the modified entropy and the attractor 
equation. Surprisingly, the results again match with the equations guessed 
in \cite{9906094,0007195}. Thus in a sense our analysis gives a derivation 
of the 
equations conjectured in \cite{9906094,0007195}, although in the context 
of a
different (but related) theory.

The rest of the paper is organized as follows. In section \ref{s1.5} we
review the construction of the black hole entropy function for an extremal
black hole in four dimensions with near horizon geometry $AdS_2\times S^2$
and use this formalism to demonstrate that the black hole entropy is
independent of the asymptotic values of the moduli scalar fields. We also
show that the entropy computed using this formalism is unchanged under a
field redefinition of the metric and scalar fields and also under an
electric-magnetic duality transformation. In section \ref{s2} we analyze
extremal black hole solutions in heterotic string theory compactified on
$\MM\times T^2$ where $\MM$ stands for a suitable compact manifold ({\it
e.g.} $T^4$ or $K3$ or some orbifolds of these.) We analyze black hole 
solutions carrying momentum and winding charges, as well as Kaluza-Klein 
monopole and H-monopole charges associated with the two circles of $T^2$, 
-- first in the supergravity approximation and then including
the Gauss-Bonnet term that arises at the heterotic string tree level. We 
find 
that the final result for the entropy as well as the near horizon values 
of the metric, gauge field strengths and the axion-dilaton field are 
identical to the ones found in 
\cite{9801081,9812082,9904005,9906094,9910179,0007195,0009234,0012232} 
based on supersymmetrized Weyl tensor squared terms. In section \ref{shol} 
we focus on the special case of $\NN=4$ supersymmetric heterotic string 
compactification, -- either by taking heterotic string theory on $T^6$ or 
by 
considering more general class of CHL 
compactifications\cite{9505054,9506048,9507027,9507050,9508144,9508154}, 
--
but go 
beyond the tree approximation and include higher order corrections so as 
to restore S-duality of the effective theory. We analyze the black hole 
entropy in the modified effective field theory using the entropy function, 
and find an S-duality covariant expression for the black hole entropy and 
the near horizon value of the axion-dilaton field. These  equations agree 
with the form of the answer guessed in \cite{9906094,0007195} using the 
requirement of 
S-duality invariance. 

Appendix \ref{sa} is devoted to fixing the normalization of 
various electric and magnetic charges which arise in this theory.

\sectiono{Black Hole Entropy Function} \label{s1.5}

Since we shall be analyzing extremal black hole solutions in four 
dimensional
string theory, 
we begin by briefly reviewing the results of \cite{0506177} in four
dimensions. Let us consider a theory of gravity coupled to a set of 
abelian gauge fields $A^{(i)}_\mu$ and
a set of neutral scalar
fields $\{\phi_s\}$, described by a gauge and general coordinate
invariant Lagrangian density $\LL$. In this theory we consider an extremal
black hole solution with near horizon geometry $AdS_2\times S^2$. The
most general near horizon field configuration consistent with the 
$SO(2,1)\times SO(3)$ symmetry of $AdS_2\times S^2$ is of the
form:
\ben \label{e0}
&&  {ds^2  = v_1\left(-r^2 dt^2+{dr^2\over 
r^2}\right)   
+
v_2 \left(d\theta^2+\sin^2\theta d\phi^2\right) }\nonumber \\
&& \phi_s =u_s \nonumber \\ 
&&  {F^{(i)}_{rt} = e_i,}  \qquad  F^{(i)}_{\theta\phi} = {p_i \over 4\pi} \,  
\sin\theta\, , 
\een
where $F^{(i)}_{\mu\nu}=\p_\mu A^{(i)}_\nu - \p_\nu A^{(i)}_\mu$ and
$v_1$,  $v_2$,
$\{u_s\}$, $\{e_i\}$ and $\{p_i\}$ are constants labelling the
background. We now define:
\be \label{e1}
f(\vec u, \vec v, \vec e, \vec p)\equiv\int d\theta \, 
d\phi\, \sqrt{-\det g} \, \LL\,
\ee
evaluated for the background \refb{e0}. Furthermore we define
\be \label{e2}
{ q_i\equiv {\p f \over \p e_i} ,} \qquad
 {F(\vec u, \vec v, \vec q, \vec p) \equiv 2\, 
\pi ( e_i \, q_i - f(\vec u, \vec v, \vec e, \vec p)) }
\ee
so that $F/2\pi$ is the Legendre transform of 
the function $f$ with respect to
the variables $\{e_i\}$.
Then it follows as a consequence of the equations of motion that for a
black hole carrying electric 
charge $\vec q$ and magnetic charge $\vec p$,
the constants
$\vec v$, $\vec u$ and $\vec e$ are given by:
\be \label{e3}
{ {\p F \over \p u_s}=0, \qquad {\p
F \over \p v_1}=0\,, \qquad  {\p
F \over \p v_2}=0}\, .
\ee
\be \label{e4}
e_i = {1\over 2\pi} \, {\p F(\vec u, \vec v, \vec q, \vec p) 
\over \p q_i}  \, .
\ee
Furthermore, using the results of 
\cite{9307038,9312023,9403028,9502009} one can prove that 
the entropy associated with the black hole is given by:
\be \label{e5}
S_{BH} = F(\vec u, \vec v, \vec q, \vec p) \, 
\ee
evaluated at the extremum \refb{e3}.

An alternative but equivalent formulation is to treat $\vec e$ and $\vec 
q$ as 
independent variables, and define:
\be \label{e2a}
 {F(\vec u, \vec v, \vec e, \vec q, \vec p) \equiv 2\, 
\pi ( e_i \, q_i - f(\vec u, \vec v, \vec e, \vec p)) }
\ee
The equations determining $\vec u$, $\vec v$ and $\vec e$ are then given 
by:
\be \label{e3a}
{ {\p F \over \p u_s}=0, \quad {\p
F \over \p v_1}=0\,, \quad  {\p
F \over \p v_2}=0}\, , \quad {\p F\over \p e_i}=0\, .
\ee
The entropy associated with the black hole is given by:
\be \label{e5a}
S_{BH} = F(\vec u, \vec v, \vec e, \vec q, \vec p) \, ,
\ee
at the extremum \refb{e3a}.

It is worth emphasizing again that these results follow as a consequence 
of the equations of motion and 
Wald's formula for entropy\cite{9307038,9312023,9403028,9502009} in the 
presence of higher derivative 
terms\cite{0506177}. Supersymmetry was not used in this analysis.
However in the special case of $\NN=2$ supersymmetric theories,
these results reproduce the observation of \cite{0405146} that the
Legendre transform of the entropy with respect to the electric charges
gives the prepotential of the theory\cite{0506177}.

Before concluding this section we shall discuss some important 
consequences 
of this result:

\begin{enumerate}

\item Since the construction of the function $F$ involves knowledge of 
only the 
Lagrangian density,  the functional form of
$F$ is independent of asymptotic values of 
the moduli scalar 
fields. Thus if the extremization equations \refb{e3} 
determine all the parameters $\vec u$, $\vec v$, then the value of $F$ at 
the extremum and hence the entropy $S_{BH}$ is completely independent of 
the asymptotic values of the moduli fields. If on the other hand the 
function $F$ has flat directions then only some combinations of
the parameters $\vec u$, $\vec v$ are
determined by extremizing $F$, and the rest
may depend on the asymptotic values of 
the moduli fields. However since $F$ is independent of the flat 
directions, it depends only on the combination of parameters which are 
fixed by the extremization equations. As a result the value of $F$ at the 
extremum is still independent 
of the asymptotic moduli. Thus the entropy of the black hole is 
independent of the asymptotic values of the moduli fields irrespective of 
whether or not $F$ has flat directions. This is a generalization of the usual
attractor mechanism for supersymmetric black holes in supergravity 
theories\cite{9508072,9602111,9602136}.

\item An arbitrary field redefinition of the metric and the scalar 
fields will induce a redefinition of the parameters $\vec u$, $\vec v$, 
and hence the functional form of $F$ will change. However, since the value 
of $F$ at the extremum is invariant under non-singular field redefinition,
the entropy is unchanged under a redefinition of the metric and 
other scalar fields. To  see this more explicitly, let us consider a 
reparametrization of $\vec u$ and $\vec v$ of the form:
\be \label{erepar}
\wh u_s = g_s(\vec u, \vec v, \vec e, \vec p), \quad 
\wh v_i = h_i(\vec u, \vec v, 
\vec e, \vec p)\, ,
\ee
for some functions $\{g_s\}, \{h_i\}$.
Then it follows from eqs.\refb{e1}, \refb{e2a} that the new entropy 
function $\wh F(\vec{\wh u}, \vec{\wh v}, \vec e, 
\vec q, \vec p)$ is given by:
\be \label{erepar1}
\wh F(\vec{\wh u}, \vec{\wh v}, \vec e, \vec q, \vec p)
= F(\vec u, \vec v, \vec e, \vec q, \vec p)\, .
\ee
It is easy to see that eqs.\refb{e3a} are equivalent to:
\be \label{e3b}
{ {\p \wh F \over \p \wh u_s}=0, \quad {\p
\wh F \over \p \wh v_1}=0\,, \quad  {\p
\wh F \over \p \wh v_2}=0}\, , \quad {\p \wh F\over \p e_i}=0\, .
\ee
Thus the value of $\wh F$ evaluated at this extremum is equal to the value 
of $F$ evaluated at the extremum \refb{e3a}.
This result of course is a consequence of the field 
redefinition invariance of Wald's entropy formula as discussed in 
\cite{9312023}.

\item As is well known, Lagrangian density is not invariant under an 
electric-magnetic duality transformation. However, the function $F$, being 
Legendre transformation of the Lagrangian density with respect to the 
electric field variables, is invariant under an electric-magnetic duality 
transformation. In other words, if instead of the original Lagrangian 
density $\LL$, we use an equivalent dual Lagrangian density
$\wt \LL$ where some 
of the gauge fields have been dualized, and construct a new entropy
function $\wt F(\vec u, \vec v, \vec q, \vec p)$ from this new Lagrangian 
density, then $F$ and $\wt F$ are related to each other by exchange of the
appropriate $q_i$'s and $p_i$'s. In the context of two derivative 
theories this point has been noted in \cite{0505122}.

\end{enumerate}

\sectiono{Dyonic Black Holes in Heterotic String Theory}
\label{s2}

We shall now apply the results described in section \ref{s1.5}
to heterotic string theory on $\MM\times S^1
\times \wt S^1$, where $\MM$ is some four manifold (possibly accompanied
by background gauge fields and anti-symmetric tensor fields) suitable for
heterotic string compactification.
Examples of $\MM$ are $K3$ or $T^4$,
but more general orbifold compactifications are also possible. We 
use $\alpha'=16$ unit and denote by $x^8$ and $x^9$ the coordinates along
$\wt S^1$ and $S^1$ respectively. We also normalize the coordinates $x^8$ and
$x^9$ such that they have periodicity $2\pi \sqrt{\alpha'}=8\pi$. We 
denote by $n$ and $w$ the number of units of momentum and winding along 
$S^1$, by $\wt n$ and $\wt w$ the number of units of momentum and winding 
along $\wt S^1$, by $N$ and $W$ the number of units of Kaluza-Klein 
monopole charge\cite{grossperry,sorkin} and $H$-monopole 
charge\cite{9211056} associated with $S^1$ and by 
$\wt N$ and $\wt W$ the number of units of Kaluza-Klein monopole charge 
and $H$-monopole charge associated with $\wt S^1$.\footnote{A Kaluza-Klein 
monopole associated with $S^1$ represents a background where the circle 
$S^1$ is non-trivially fibered over the two sphere labelled by 
$\theta,\phi$. An $H$-monopole associated with $S^1$ represents a 
five-brane wrapped on $\MM\times\wt S^1$.}

Although eventually we shall be interested in studying a general black 
hole solution carrying all the eight charges, we shall first consider black 
hole solution with non-zero $n$, $w$ and $\wt N$, $\wt W$, setting $\wt 
n=\wt w=N=W=0$. In the supergravity approximation the four dimensional 
fields relevant for the construction 
of this black hole solution
are related to the ten dimensional string metric $G^{(10)}_{MN}$,
anti-symmetric tensor field $B^{(10)}_{MN}$ and the dilaton $\Phi^{(10)}$ 
via the 
relations:\footnote{We use the symmetry $x^8\to -x^8$ together with a 
parity transformation of the non-compact directions to set 
$G_{89}=0$, $B_{89}=0$ and $B_{\mu\nu}=0$.}
\ben \label{e6}
 &&  {\Phi = \Phi^{(10)} - {1\over 4} \, \ln (G^{(10)}_{99})}
 - {1\over 4} \, \ln (G^{(10)}_{88}) - {1\over 2}\ln V_\MM
 \, ,\nonumber \\
&&  {S=e^{-2\Phi}}\, , \qquad
  {R = \sqrt{G^{(10)}_{99}}}\, , \qquad
 \wt R = \sqrt{G^{(10)}_{88}}\, , \nonumber \\ 
&& {G_{\mu\nu} = G^{(10)}_{\mu\nu} - (G^{(10)}_{99})^{-1} \,
G^{(10)}_{9\mu}
\, G^{(10)}_{9\nu} - (G^{(10)}_{88})^{-1} \,
G^{(10)}_{8\mu}
\, G^{(10)}_{8\nu}\, , }\nonumber \\ 
&& A^{(1)}_\mu = {1\over 2} (G^{(10)}_{99})^{-1} \, G^{(10)}_{9\mu}\, ,
\qquad   {A^{(2)}_\mu = {1\over 2} (G^{(10)}_{88})^{-1} \, G^{(10)}_{8\mu}\, ,}
\nonumber \\ 
&&  {A^{(3)}_\mu = {1\over 2} B^{(10)}_{9\mu}\, , }
\qquad {A^{(4)}_\mu = {1\over 2} B^{(10)}_{8\mu}}\, ,
 \een
 where $V_\MM$ denotes the volume of $\MM$ measured in the string
 metric.
 The effective action involving these fields is given by
 \ben \label{e4+}
&&\SSS = {1\over 32\pi} \int d^4 x \, \sqrt{-\det G} \,
S \, \bigg[ R_G 
+ S^{-2}\, G^{\mu\nu} \, \p_\mu S \p_\nu S \nonumber \\
&& \quad -  R^{-2}
\, G^{\mu\nu} \, \p_\mu R \p_\nu R -  \wt R^{-2}
\, G^{\mu\nu} \, \p_\mu \wt R \p_\nu \wt R \nonumber \\
&& \, - R^2 \, 
G^{\mu\nu} \, G^{\mu'\nu'} \, F^{(1)}_{\mu\mu'} 
F^{(1)}_{\nu\nu'} - \wt R^2 \, 
G^{\mu\nu} \, G^{\mu'\nu'} \, F^{(2)}_{\mu\mu'} 
F^{(2)}_{\nu\nu'}
 \nonumber \\
&& \, - R^{-2} \,
G^{\mu\nu} \, G^{\mu'\nu'} \, F^{(3)}_{\mu\mu'}
F^{(3)}_{\nu\nu'} 
 - \wt R^{-2} \,
G^{\mu\nu} \, G^{\mu'\nu'} \, F^{(4)}_{\mu\mu'}
F^{(4)}_{\nu\nu'}\bigg] \nonumber \\
&& \qquad +\hbox{ higher derivative terms 
+ string loop corrections}
\een
{}From the definition of $A^{(i)}_\mu$ given in \refb{e6} it follows that the
fields $A^{(1)}_\mu$ and $A^{(3)}_\mu$ couple to the momentum and
winding numbers along the $x^9$ direction, whereas the fields 
$A^{(2)}_\mu$ and $A^{(4)}_\mu$ couple to the momentum and winding
numbers along the $x^8$ direction. Thus the electric charges $q_1$ and
$q_3$ associated with the fields $A^{(1)}_\mu$ and $A^{(3)}_\mu$ are
proportional to $n$ and $w$ respectively, while the magnetic charges
$p_2$ and $p_4$ associated with the fields $A^{(2)}_\mu$ and
$A^{(4)}_\mu$ are proportional to the quantum numbers $\wt N$ and
$\wt W$ respectively. The constants of proportionality have been determined
in appendix \ref{sa} and the result is as follows:
\be \label{erel}
q_1={1\over 2} n, \quad q_3={1\over 2} w, \quad p_2=4\pi\wt N, \quad
p_4=4\pi \wt W\, .
\ee

We now consider an extremal black hole solution in this theory with near
horizon geometry:
\ben \label{eb1}
 {ds^2  = v_1\left(-r^2 dt^2+{dr^2\over 
r^2}\right)   
+
v_2 (d\theta^2 + \sin^2\theta d\phi^2)\, , }\nonumber \\ 
 S =u_S, \qquad R=u_R , \qquad  {\wt R = u_{\wt R} }\nonumber \\
  {F^{(1)}_{rt} = e_1,} \quad F^{(3)}_{rt}=e_3, \quad
 {F^{(2)}_{\theta\phi}={p_2\over 4\pi}, }\quad F^{(4)}_{\theta\phi}
= {p_4\over 4\pi}    
\een
For this background geometry, we shall first compute the function $f$
defined in \refb{e1} by ignoring the higher derivative terms and string loop
corrections in the action
\refb{e4+}. 
A straightforward calculation gives:
\ben \label{ey2}
&& f(u_S, u_R,\wt u_R, v_1,v_2, e_1, e_3,p_2, p_4)   
{\equiv \int d\theta d\phi \, \sqrt{-\det G} \, \LL}
\nonumber \\ 
&=& 
{1\over 8} \, v_1 \, v_2
\, u_S  \left[ -{2\over v_1} +{2\over v_2} + 
{2 \,  u_R^2 \, e_1^2\over v_1^2} +{2 \, e_3^2 \over u_R^2 \, v_1^2}
 - 2 \, u_{\wt R}^2 \, {p_2^2 \over 16\pi^2 v_2^2} - 2\,
 u_{\wt R}^{-2} \,
{p_4^2 \over 16\pi^2 v_2^2}
\right] \nonumber \\
\een
Eqs.\refb{e2} now give:
\be \label{eb2}
q_1= {1\over 2}\, u_S\, {v_2\over v_1} \, u_R^2\, \, e_1, \quad 
q_3= {1\over 2}\, u_S\, {v_2\over v_1} \, u_R^{-2}\, e_3\, ,
\ee
and
\ben \label{eb3}
&& F(u_S, u_R,\wt u_R, v_1,v_2, q_1, q_3,p_2, p_4) \nonumber \\ 
&=& 
{\pi\over 4} \, v_1 \, v_2
\, u_S  \left[ {2\over v_1} - {2\over v_2} + 
{8 \,   q_1^2\over u_R^2\, v_2^2\, u_S^2} +{8 \, u_R^2\, 
q_3^2 \over  v_2^2\, u_S^2}
\right.  
\left. + 2 \, u_{\wt R}^2 \, {p_2^2 \over 16\pi^2 v_2^2} + 2\,
 u_{\wt R}^{-2} \,
{p_4^2 \over 16\pi^2 v_2^2}
\right]
\nonumber \\
\een
It is now straightforward to solve eqs.\refb{e3}. The result 
is:\footnote{We are implicitly assuming that $q_i$ and $p_i$ are all 
positive. Otherwise $q_i$, $p_i$ must be replaced by $|q_i|$, $|p_i|$ in 
the final formul\ae. \label{f3}} 
\ben 
\label{eb4}
&& v_1=v_2 = {1\over 4\pi^2} p_2 p_4 = 4 \wt N\, \wt W \nonumber \\ 
&& {u_S = 8\pi \, \sqrt{q_1 q_3\over p_2 p_4} = 
\sqrt{n\, w\over \wt N\, \wt W} \ },
\quad 
u_R = \sqrt{q_1\over q_3}=\sqrt{n\over w}, \quad  {u_{\wt R}
=\sqrt{p_4\over p_2} = \sqrt{\wt W\over \wt N}\, .}\nonumber \\
\een
Eq.\refb{e5}  now  gives
\be \label{eb5} 
{S_{BH} = \sqrt{q_1q_3p_2 p_4} = 2\pi \sqrt{nw\wt N\wt W} }.
\ee
This agrees with the standard result for the 
entropy of four charge black holes
in (3+1) dimensions.

By examining the background \refb{eb4} we see that the effective string
coupling square at the horizon is given by  
$u_S^{-1}\sim \sqrt{\wt N\wt W/nw}$, whereas
the sizes of $AdS_2$ and $S^2$, measured in string metric, are of order
$\sqrt{v_1}, \sqrt{v_2}\sim \sqrt{\wt N \wt W}$. Finally the squares of various
field strengths appearing in the action are of order ${1/\wt N\wt W}$. This
shows that in this background 
the string loop expansion is controlled by the combination 
$\sqrt{\wt N\wt W/ nw}$ and the $\alpha'$ expansion is 
controlled by the parameter
${1/\wt N\wt W}$. From now on we shall work in the limit $nw>>\wt N\wt W$
so that higher loop correction terms are negligible, and work with the
tree level heterotic string theory. In this case we expect that the $\alpha'$
corrections to the solution and the entropy will generate a power series
expansion in $1/\wt N\wt W$.

We shall now consider a specific higher derivative correction to the action,
namely the Gauss-Bonnet term.\footnote{There is no {\it a priori} reason to
believe that this calculation would reproduce the correct entropy for
heterotic black holes even for large $nw/\wt N \wt W$, since the full tree
level effective action contains other terms. However we shall see that the
final result for the entropy agrees with that of Cardoso {\it et.al.}, which
in turn is known to reproduce correctly the microscopic entropy of the
black hole.}
At tree level in
heterotic string theory this corresponds to an additional term in
 the Lagrangian density of the 
form\cite{rzwiebach}\footnote{The Lagrangian density also contains a term
involving the product of the axion field $a$ and the Pontryagin 
density\cite{9610237}, but
this term vanishes identically in 
$AdS_2\times S^2$ background, and we do not
need to include this term in our analysis. \label{f2}}$^,$\footnote{Some
recent discussion on the effect of Gauss-Bonnet and 
other higher derivative terms
on black hole solutions 
can be found in \cite{0112045,0202140,0212092,0302136,0408200}.}
\be \label{ec1}
\Delta\LL =  {S\over 16\pi}\, 
\left\{ R_{G\mu\nu\rho\sigma} R_G^{\mu\nu\rho\sigma} 
- 4 R_{G\mu\nu} R_G^{\mu\nu}
+ R_G^2
\right\} \, ,
\ee
where $R_{G\mu\nu\rho\sigma}$ denotes the Riemann tensor computed using
the string metric $G_{\mu\nu}$.
This induces the following change in the functions $f$ and $F$
\be \label{ec2}
 \Delta f = -2 \, u_S \, \quad 
\to \quad {\Delta F = 4\, \pi \, u_S}
\ee
and does not change the relation \refb{eb2} 
between $q_i$ and $e_i$, since the
correction term is independent of $e_i$. Eqs.\refb{e3}, \refb{e5}
with the new function $F+\Delta F$ now give:
\ben \label{eb4n}
&& v_1=v_2  = 4\, \wt N\, \wt W +8 \nonumber \\ 
&& u_S =  \sqrt{n\, w\over \wt N\, \wt W+4}  \, ,
\quad 
u_R =  \sqrt{n\over w}, \quad  {u_{\wt R}
 = \sqrt{\wt W\over \wt N}\, .}\nonumber \\
\een
and
\be \label{ed1} 
S_{BH}  = 
2\pi \, \sqrt{nw} \sqrt{\wt N\wt W + 4} \, .
\ee

For $\wt N=\wt W=0$ in \refb{ed1} we get 
$S_{BH}=4\pi\sqrt{nw}$. This agrees with the results of 
\cite{0409148,0410076,0411255,0411272,0501014} (which in turn agrees 
with the microscopic counting based on degeneracy of elementary string 
states) even though the action 
used in the analysis of these papers was quite different. This surprising 
agreement was noted 
in \cite{0505122} and will be extended and discussed in more detail later 
in this 
section.
We also note that if we set $\wt N=\wt W=0$ and take $nw$ to be
negative, we get $S_{BH}=4\pi\sqrt{|nw|}$ (see footnote \ref{f3}). 
On the other hand these
configurations correspond to non-BPS fundamental heterotic string states
with only right-moving excitations on the world-sheet. The expected
microscopic entropy for this system is $2\sqrt 2\pi\sqrt{nw}$. Thus the
Gauss-Bonnet correction to the low energy effective action is not able to
reproduce the microscopic entropy of these non-BPS states.

We shall now extend our analysis to a general black hole solution 
carrying all eight charges $n$, $w$, $\wt n$, $\wt w$, $N$, $W$, $\wt N$, 
$\wt W$. In order to describe 
a black hole configuration of this type, we need to include in our 
analysis a more general set of fields. These include the metric 
$G_{\mu\nu}$, four gauge fields $A^{(i)}_\mu$ ($1\le i\le 4$), the axion 
field $a$, the dilaton field $S$,   
and a real $4\times 4$ matrix valued scalar field $M$ satisfying:
\be \label{eag1}
MLM^T=M, \quad M^T=M, \quad L\equiv \pmatrix{0 & 0 & 1& 0 \cr 0 & 0 & 0 & 
1 \cr 1 & 0 & 0 & 0\cr 0 & 1 & 0 & 0}\, .
\ee
In the supergravity approximation these fields are related to the ten 
dimensional fields $G^{(10)}_{MN}$, 
$B^{(10)}_{MN}$ and $\Phi^{(10)}$ as follows (see {\it e.g.} \cite{9402002}):
\ben \label{eag2}
&& \wh G_{mn} = G^{(10)}_{mn}, \quad \wh B_{mn} = B^{(10)}_{mn}\, , 
\qquad 
m,n=8,9\, , 
\nonumber 
\\  && \wh G^{mn} = (\wh G^{-1})^{mn} \, , \nonumber  \\
&& M =  
\pmatrix{ \wh G^{-1} & \wh G^{-1} B \cr -\wh B \wh G^{-1} & \wh 
G - \wh B \wh G^{-1} \wh B} \nonumber \\
&& A^{(10-m)}_\mu = {1\over 2} \wh G^{mn} G^{(10)}_{m\mu} , \quad
A^{(12-m)}_\mu = {1\over 2} B^{(10)}_{m\mu} - \wh B_{mn} A^{(10-m)}_\mu, \qquad
0\le \mu, \nu \le 3 \, , \nonumber \\
&& G_{\mu\nu} = G^{(10)}_{\mu\nu} - \wh G^{mn} G^{(10)}_{m\mu} 
G^{(10)}_{n\nu}\, , \nonumber \\
&& B_{\mu\nu} = B^{(10)}_{\mu\nu} 
- 4 \wh B_{mn} A^{(10-m)}_\mu A^{(10-n)}_\nu - 2 (A^{(10-m)}_\mu A^{(12-m)}_\nu
- A^{(10-m)}_\nu A^{(12-m)}_\mu) \, , \nonumber \\
&& \Phi = \Phi^{(10)} - {1\over 4} \ln \det \wh G -{1\over 2} \ln V_\MM
\, , \qquad S = e^{-2\Phi}\, . 
\een
In interpreting $\wh G$ and $\wh B$ 
as matrices we must take $m=9$
($n=9$) as a label of the first row (column) 
and $m=8$ ($n=8$) as a label of the second row (column). Thus
for example $\wh G_{99}$ appears in the top left hand corner
of the matrix $\wh G$.
Finally $a$ is defined through the relation
\ben \label{eag2a}
H_{\mu\nu\rho} &\equiv& (\p_\mu B_{\nu\rho} + 2 A_\mu^{(i)} L_{ij} 
F^{(j)}_{\nu\rho}) + \hbox{cyclic permutations of $\mu$, $\nu$, $\rho$}\, , 
\qquad 1\le i,j\le 4\, , \nonumber \\
&=& G_{\mu\mu'} G_{\nu\nu'}G_{\rho\rho'} S^{-1} \, (\sqrt{-\det G})^{-1}
\, \epsilon^{\mu'\nu'\rho'\sigma'} \, \p_{\sigma'} a \, .
\een
In terms of these four dimensional fields, 
the effective action in the supergravity approximation is given by:
\ben \label{eag3}
\SSS &=& {1\over 32\pi}\, \int d^4 x \, \sqrt{-\det G} \,  S\, 
\left[ R_G + {1\over S^2}\,
G^{\mu\nu}  (\p_\mu S \p_\nu S -{1\over 2} 
\p_\mu a \p_\nu a) +{1\over 8}
G^{\mu\nu} Tr(\p_\mu M L \p_\nu M L) \right. \nonumber \\
&& \left. -  G^{\mu\mu'} G^{\nu\nu'}\,
F^{(i)}_{\mu\nu} (LML)_{ij} F^{(j)}_{\mu'\nu'}  - {a\over S}
G^{\mu\mu'} G^{\nu\nu'}\,
F^{(i)}_{\mu\nu} L_{ij} \wt F^{(j)}_{\mu'\nu'}  \right]\, .
\een

We now look for a near horizon field configuration of the form:
\ben \label{ehor}
 {ds^2  = v_1\left(-r^2 dt^2+{dr^2\over 
r^2}\right)   
+
v_2 (d\theta^2+\sin^2\theta\, d\phi^2) \, , }\nonumber \\ 
 S =u_S, \qquad a=u_a , \qquad  {M_{ij} = u_{Mij} }\nonumber \\
  {F^{(i)}_{rt} = e_i,} \quad  
 {F^{(i)}_{\theta\phi}={p_i\over 4\pi}} \, .  
 \een
Substituting \refb{ehor} into \refb{eag3} and using \refb{e1} we get
\ben \label{eag5}
&& f(u_S, u_a, u_M, \vec v, \vec e, \vec p)
\equiv \int d\theta d\phi \, \sqrt{-\det G} \, \LL
\nonumber \\ 
&=& 
{1\over 8} \, v_1 \, v_2
\, u_S  \left[ -{2\over v_1} +{2\over v_2} + 
 {2\over v_1^2} e_i (Lu_M L)_{ij} e_j - {1\over 8\pi^2 v_2^2}  p_i 
 (Lu_ML)_{ij} p_j + { u_a\over \pi u_S v_1 v_2}  e_i L_{ij} p_j
 \right] \nonumber \\
\een
Its Legendre transform with respect to the variables $e_i$ gives the entropy
function $F$:
\be \label{eag6}
q_i \equiv {\p f\over \p e_i} = {v_2 u_S\over 2 v_1} (Lu_M L)_{ij} e_j 
+ {u_a\over 8\pi }   L_{ij} p_j\, ,
\ee
\ben \label{eag7}
F(u_S, u_a, u_M, \vec v, \vec q, \vec p) &\equiv& 2\pi \left( e_i q_i 
- f(u_S, u_a, u_M, \vec v, \vec e, \vec p) \right) \nonumber \\
&=& 2\pi \bigg[ {u_S\over 4} (v_2 - v_1)
+{v_1\over v_2 u_S} \, q^T u_M q 
+{v_1\over 64\pi^2 v_2 u_S} (u_S^2 + u_a^2)
p^T L u_M L p \nonumber \\
&& -{v_1\over 4 \pi v_2 u_S}  \, u_a\, q^T u_M L p\bigg]\,.
\een
As shown in appendix \ref{sa}, the charges $\vec q$, $\vec p$ 
are related to the quantum numbers $n$, $w$,
$\wt n$, $\wt w$, $N$, $W$, $\wt N$, $\wt W$ as
\ben \label{eag8}
&&
q_1={1\over 2}\, n, \quad q_2={1\over 2}\, \wt n, \quad q_3=
{1\over 2}\, w, \quad q_4={1\over 2}\, 
\wt w\, , \nonumber \\
&&
p_1=4\pi N, \quad p_2=4\pi \wt N, \quad p_3=4\pi W, \quad p_4=4\pi\wt W\, .
\een
This suggests that we define new charge vectors:
\be \label{eag8a}
Q_i = 2 q_i, \qquad P_i = {1\over 4\pi}\, L_{ij} p_j,
\ee
so that $P_i$ and $Q_i$ are integers. In terms of $\vec Q$ and $\vec P$
the entropy function is given by:\footnote{As expected, $F$ is invariant 
under the 
S-duality transformation $\pmatrix{Q'\cr P'}=\pmatrix{ m & 
n\cr r & s} \pmatrix{Q\cr P}$, $u_a'+ i u_S' = \{m (u_a + i u_S) + n\} / 
\{ r (u_a + i u_S) + s)\}$, $v_i'=  v_i u_S /u_S'$. \label{f1}}
\be \label{eag8b}
F = {\pi\over 2} \bigg[ u_S(v_2 - v_1) +{v_1\over v_2 u_S} \left( Q^T u_M 
Q +
 (u_S^2 + u_a^2) \, P^T u_M P - {2 }
\, u_a \, Q^T u_M P \right)\bigg]\, .
\ee

We now need to find the extremum of $F$ with respect to $u_S$, $u_a$,
$u_{Mij}$, $v_1$ and $v_2$. In general this leads to a complicated set
of equations. However we can simplify the analysis by noting that the action
\refb{eag3} has an
$SO(2,2)$ symmetry acting
on $M$ and $F^{(i)}_{\mu\nu}$:
\be \label{eag8c}
M\to \Omega M \Omega^T, \quad F^{(i)}_{\mu\nu}\to \Omega_{ij} 
F^{(j)}_{\mu\nu}\, ,
\ee   
where $\Omega$ is a matrix satisfying
\be \label{eag9}
\Omega^T L \Omega=L\, .
\ee
\refb{eag8c} induces the following transformation on the various parameters:
\ben \label{eag9a}
e_i\to \Omega_{ij} e_j, \qquad  p_i\to \Omega_{ij} p_j, \qquad u_M\to
\Omega u_M \Omega^T\, , \nonumber \\
q_i\to (L\Omega L)_{ij} q_j \, , \qquad
Q_i\to (L\Omega L)_{ij} Q_j, \qquad P_i \to (L\Omega L)_{ij} P_j\, .
\een
The entropy function \refb{eag8b} is invariant under these 
transformations.
Since at its extremum with respect to $u_{Mij}$
the entropy function depends only on $\vec P$, $\vec Q$, $v_1$, $v_2$, 
$u_S$ and $u_a$ it must be a function
of  the $SO(2,2)$ invariant combinations:
\be \label{eag10}
Q^2 = Q_i L_{ij} Q_j, \quad P^2 = P_i L_{ij} P_j, \quad 
Q\cdot P = Q_i L_{ij} P_j\, ,
\ee
besides $v_1$, $v_2$,
$u_S$ and $u_a$.
Let us for definiteness take $Q^2>0$, $P^2>0$, 
and $(Q\cdot P)^2<Q^2 P^2$. In that case with the help
of an $SO(2,2)$ transformation we can make 
\be \label{eag11}
(I-L)_{ij}Q_j=0, \quad (I-L)_{ij}P_j=0\, ,
\ee
where $I$ denotes the $4\times 4$ identity matrix.
It can be easily seen that for $\vec P$ and 
$\vec Q$ satisfying this condition, every
term in \refb{eag8b} is extremized 
with respect to $u_{M}$ for\footnote{This
is most easily seen by diagonalizing 
$L$ to the form $\pmatrix{I_2 & \cr & -I_2}$
where $I_2$ is a $2\times 2$ identity 
matrix. In this case $Q$ and $P$ satisfying \refb{eag11}
will have
third and fourth components zero, and hence a variation $\delta u_{Mij}$
with either $i$ or $j$ = 3, 4 will give vanishing contribution to each term in
$\delta F$ computed from \refb{eag8b}. 
On the other hand due to the constraint \refb{eag1} on $M$, any 
variation $\delta M_{ij}$ (and hence 
$\delta u_{Mij}$) with $i,j=1,2$ must vanish.
Thus each term in $\delta F$ vanishes under all the 
allowed variations of $u_M$.}
\be \label{eag12}
u_M  = I\, .
\ee
Substituting \refb{eag12} into \refb{eag8b} and using \refb{eag10},
\refb{eag11}, we get:
\be \label{eag13}
F= {\pi\over 2} \left[ u_S(v_2-v_1) 
+{v_1\over v_2} \left({Q^2\over u_S} + 
{P^2\over u_S} (u_S^2 + u_a^2)
- 2\, {u_a\over u_S}\, Q\cdot P \right) \right]\, .
\ee
Written in this $SO(2,2)$ invariant manner, eq.\refb{eag13} is valid
for general $\vec P$, $\vec Q$ satisfying $P^2>0$, $Q^2>0$ and 
$(Q\cdot P)^2<Q^2P^2$.
 
It remains to extremize $F$ with respect to $v_1$, $v_2$, $u_S$ and $u_a$.
This gives
\be \label{eag14}
v_1=v_2= 2 P^2 \, ,
\qquad u_S= {\sqrt{Q^2 P^2 - (Q\cdot P)^2}\over P^2} \, ,
\qquad u_a = 
{Q\cdot P\over P^2}\, .
\ee
The black hole entropy, given by the value of $F$ for this configuration,
is 
\be \label{eag15}
S_{BH} = \pi \, \sqrt{Q^2 P^2 - (Q\cdot P)^2}\, .
\ee

Finally let us discuss the effect of adding the Gauss-Bonnet term given in
\refb{ec1}. Since this does not affect the relation between $\vec q$ and 
$\vec e$,
we get, as in \refb{ec2},
\be \label{ec2a}
 \Delta f = -2 \, u_S \, \quad 
\to \quad {\Delta F = 4\, \pi \, u_S}\, .
\ee
For $F$ given in \refb{eag8b} and $Q$, $P$ satisfying \refb{eag11}  we 
have an extremum of $F+\Delta F$ 
at
\be \label{eum}
u_M=I,
\ee
\ben \label{eag16}
&& u_S= \sqrt{Q^2 P^2 - (Q\cdot P)^2\over P^2 (P^2+8)} \, ,
\qquad u_a = 
{Q\cdot P\over P^2}\, ,
\nonumber \\
&& v_1=v_2 =  2 P^2+8\, , 
\een
and
\be \label{eag17}
S_{BH}=F = \pi \sqrt{ (P^2+8) (Q^2 P^2 - (Q\cdot P)^2)\over
P^2}\, .
\ee
Since eqs.\refb{eag16}, \refb{eag17} 
are written 
in an $SO(2,2)$ covariant form, they are valid for general $\vec Q$, $\vec 
P$ 
with $Q^2>0$, $P^2>0$, $(Q\cdot P)^2<Q^2 P^2$.

Another quantity of interest is the apparent entropy 
of the black hole that we would
get had we used the original Bekenstein-Hawking 
formula for our computation. Although this is not the real entropy, this
gives a convention independent measure of the near horizon metric. For the
normalization of the action used in \refb{eag3} the Newton's constant
$G_N=2$. Since the canonical metric is given by 
$g_{\mu\nu}=S \, G_{\mu\nu}$, the area of the horizon measured in the
canonical metric is $4\pi \, u_S \, v_2$. Thus the apparent entropy
is given by:
\be \label{eapp1}
S_{app}={4\pi u_S v_2\over 4 G_N} = \pi \, (P^2+4)\, 
\sqrt{Q^2 P^2 - (Q\cdot P)^2 \over P^2 (P^2+8)}\, .
\ee
Thus
\be \label{eapp2}
{S_{app}\over S_{BH}} = {P^2+4\over P^2+8}\, .
\ee

We now compare these results with the computations of 
\cite{9801081,9812082,9904005,9906094,9910179,0007195,0009234,0012232}. In 
these papers a
general method for analyzing supersymmetric black holes in the presence of 
curvature squared terms was developed. The analysis uses a fully
supersymmetric action\cite{9602060,9603191}, and the curvature squared 
term takes the form of 
Weyl tensor squared rather than the 
Gauss-Bonnet combination. The general results 
derived in these papers can be easily applied to supersymmetric black 
holes in heterotic string theory on $T^6$, $K3\times T^2$ or related 
orbifold models. Surprisingly, the result agrees exactly with the formula 
\refb{eag17} for the black hole entropy (see {\it e.g.} eq.(6.64) 
of \cite{0007195}). In fact not only the entropy but also the values of the 
dilaton-axion field given in \refb{eag16}  agree with eq.(6.63) of 
\cite{0007195}. 
Recalling that $S_{app}=\pi |Z|^2$ in the convention of \cite{0007195}
one can easily check that the ratio $S_{app}/S_{BH}$ given in \refb{eapp2}
also agrees with the results of \cite{0007195}. This in turn shows that 
the near
horizon metric is also identical in the two formalisms. Finally, since 
$S_{BH}(\vec q, \vec p)$ in the two formalisms are identical, the near 
horizon electric fields $e_i=(2\pi)^{-1}
\p S_{BH}/\p q_i$ are also 
identical.\footnote{The near horizon magnetic field is 
directly given by $p_i/4\pi$ in both formalisms.} 
Given that the starting points are quite different, -- 
we have added a Gauss-Bonnet term to 
the action without supersymmetrizing 
it, whereas ref.\cite{0007195} uses a fully supersymmetrized version of 
the 
Weyl-tensor squared term, -- this agreement is quite surprising. Perhaps 
this indicates that there is an alternative formulation of the fully 
supersymmetric action based on the Gauss-Bonnet combination that is 
simpler than the one based on the Weyl tensor squared term.

\sectiono{$N=4$ Supersymmetric Theories and Holomorphic Anomaly} 
\label{shol}

Toroidally compactified heterotic string theory
is 
S-duality invariant under 
the transformation
\be \label{es1}
\tau \to {m\tau + n \over r\tau+s}, \qquad \tau\equiv a + i S, 
\qquad m, n, r, s\in \ZZZ , \quad ms-nr=1\, ,
\ee
together with appropriate transformation on other fields\cite{9402002} 
(see footnote \ref{f1} for the action of S-duality transformation on various
parameters). The supergravity theory described by the action 
\refb{eag3} reflects this symmetry.\footnote{Although 
both for the toroidal and the CHL 
compactifications
the 
ranks of the matrices $M$ and $L$ are larger, using the continuous 
T-duality symmetry of the 
action we can align the charges of the black hole so that only 
the fields which appear in the action \refb{eag3}
are excited.}  
However the additional term \refb{ec1} 
does not respect this symmetry and hence the effective action
must receive additional corrections. 
In particular we expect the full effective Lagrangian density to contain a 
term of the 
form:\footnote{The metric that remains invariant under an 
S-duality transformation is the Einstein metric $g_{\mu\nu}=S 
G_{\mu\nu}$ and not the string 
metric. Thus in order to get a fully S-duality invaraint combination we 
need to replace the curvature tensor $R_{G\mu\nu\rho\sigma}$ in 
\refb{es2} by the curvature tensor $R_{g\mu\nu\rho\sigma}$
computed 
using the Einstein metric, and also the $\sqrt{-\det G}$ multiplying
this term in the expression for the action by $\sqrt{-\det g}$.
However  this difference is 
irrelevant  for a constant dilaton
background considered in \refb{ehor}.}$^,$\footnote{Besides 
the Gauss-Bonnet combination the action is also expected to contain a
term proportional to the Pontryagin density\cite{9610237}. 
However as discussed in
footnote \ref{f2}, contribution from this term 
vanishes for the near horizon
$AdS_2\times S^2$ geometry, and hence 
we shall not need to include this
term in our analysis.}
\be \label{es2}
\Delta\LL =  \phi(a, S)\,
\left\{ R_{G\mu\nu\rho\sigma} R_G^{\mu\nu\rho\sigma}
- 4 R_{G\mu\nu} R_G^{\mu\nu}
+ R_G^2
\right\} \, ,
\ee
where $\phi(a,S)$ is invariant under the S-duality transformation 
\refb{es1} and for weak coupling, \i.e.\ for large $S$, $\phi$ 
approaches $S/16\pi$.
For heterotic string theory on $T^6$ the correct choice of 
$\phi(a,S)$ is\cite{0007195}:
\be \label{es4}
\phi(a,S) = -{3\over 16\pi^2} \ln \left(2 \, S\, |\eta(a+iS)|^4\right)\, , 
\qquad 
\eta(\tau) \equiv e^{2\pi i\tau/24} \prod_{n=1}^\infty (1 - e^{2\pi i 
n\tau}) \, .
\ee
The form of the action \refb{eag3}, \refb{es2} is valid also for a
general four dimensional heterotic string compactification with $\NN=4$ 
supersymmetry\cite{9708062,0502126}, {\it e.g.} in CHL 
models\cite{9505054,9506048,9507027,9507050,9508144,
9508154},  with a different choice of $\phi(a,S)$.
In general $\phi(a,S)$ is of the 
form\cite{9708062,0502126}:
\be \label{es6}
\phi(a,S) = -{3\over 16\pi^2} \left[
{r-4\over 24} \, \ln \left(2 \, S\, 
\left|\eta(a+iS)\right|^4\right) 
+ \psi(a+iS) + \psi(a+iS)^*\right]\, ,
\ee
where $r$ is the total number of U(1) gauge fields in the theory and
$\psi(\tau)$ is a complex analytic function of $\tau$ in the upper half
plane such that $\p_\tau\psi(\tau)$ is a modular 
form of weight two under the S-duality group of the theory. 
$^*$ denotes complex conjugation. 
$\psi(\tau)$ itself gets shifted by constants under various modular
transformations, and  grows
linearly with $\tau$ for large $S$:
\be \label{epsitau}
\psi(\tau) \simeq {i\pi \over 144} \, (28-r)\, \tau  \, ,
\ee
so that $\phi(a,S)\simeq S/16\pi$ for large $S$.  For the $Z_N$
orbifold models discussed in \cite{0502126}:
\be \label{eadd1}
\psi(\tau) = {1\over N-1} {28-r\over 12} \left( \ln \eta(N\tau)
-\ln \eta(\tau)\right)\, .
\ee

It is now easy to study the effect of the term \refb{es2} on the entropy 
function. It gives an additional contribution to $f$ and $F$ of the form
\be \label{es7}
\Delta f = -32\pi \phi(u_a, u_S) \quad \to \quad \Delta F = 64\pi^2 
\phi(u_a, u_S)\, .
\ee
Together with \refb{eag8b} this gives
\ben \label{es8}
F &=& {\pi\over 2} \bigg[ u_S(v_2 - v_1) +{v_1\over v_2 u_S} \left( Q^T 
u_M Q +
(u_S^2 + u_a^2) \, P^T u_M P \right. \nonumber \\
&& \left. - {2 } 
\, u_a \, Q^T u_M P \right) + 128\, \pi \, \phi(u_a, u_S)\bigg]\, .
\een
As in section \ref{s2}, we can use the SO(2,2) symmetry of the 
action to align the vectors $Q$ and $P$ to be 
annihilated by $(I-L)$. In this case each term in the 
expression for $F$ is extremized 
for $u_M=I$. Substituting this back into the expression for $F$ we get
\be \label{es9}
F= {\pi\over 2} \left[ u_S(v_2-v_1) +{v_1\over v_2} \left({Q^2\over u_S} +
{P^2\over u_S} (u_S^2 + u_a^2)- 2\, {u_a\over u_S}\, Q\cdot P \right)
+ 128 \pi \phi(u_a, u_S) \right]\, .
\ee
Extremization with respect to $v_1$ and $v_2$ give:
\be \label{es10}
v_1=v_2 = u_S^{-2} \, \left({Q^2} +
{P^2} (u_S^2 + u_a^2) - 2 u_a\, Q\cdot P\right)\, .
\ee
Substituting this into \refb{es9} gives:
\be \label{es11}
F= {\pi\over 2} \bigg[ {1\over u_S}\, \left({Q^2} - 2\, {u_a}\, Q\cdot P
+
{P^2} (u_S^2 + u_a^2)\right) 
+ 128 \pi \phi(u_a, u_S)\bigg]\, .
\ee
The values of $u_a$ and $u_S$ at the horizon are determined by extremizing 
$F$ with respect to $u_a$ and $u_S$. This gives:
\ben \label{es12}
P^2 u_a - Q\cdot P + 64 \, \pi \, u_S \, {\p\phi\over \p u_a} = 0 \, ,
\nonumber \\
-{1\over u_S^2}  \left({Q^2} - 2\, {u_a}\, Q\cdot P
+
{P^2} u_a^2\right) + P^2  + 128 \, \pi \, {\p\phi\over \p u_S} = 0\, .
\een

We can now try to compare eqs.\refb{es11}, \refb{es12} with the 
corresponding results in the supersymmetric version of the theory.
Since $\phi(a,S)$ given in \refb{es4} (or more generally \refb{es6}) is 
not a sum of a holomorphic and an anti-holomorphic term, it is in general 
difficult to supersymmetrize 
the corresponding Weyl tensor squared term. Due to this reason the analysis of 
Cardoso et. al. cannot be directly extended to this case. Nevertheless 
\cite{9906094} guessed a form of the corrected equations based on the 
requirement of S-duality invariance of the theory. It can easily be shown 
that the expression 
\refb{es11} for the black hole entropy as well as the 
attractor equations \refb{es12} agree with the results guessed by Cardoso 
et. al. (see  {\it e.g.} eqs.(1) and (2) of \cite{0412287} for toroidal 
compactification). This again is a surprising agreement whose significance
needs to be explored.

{\bf Acknowledgement}:  A preliminary version 
of this work was reported in
the Strings 2005 conference. I wish to thank the 
organisers of the conference
for a stimulating environment. I would also like to thank 
M.~Shigemori for
discussion.
 
 \appendix
 
 \sectiono{Normalization of the Charges} \label{sa}
 
 In this appendix we shall fix the normalization of the electric 
 charges 
 $q_i$ and magnetic charges $p_i$ associated with the gauge
 fields $A_\mu^{(i)}$. For this we start by noting that according
 to our convention, the presence of an electric charge $q_i$
 induces a coupling
 \be \label{eapa1}
 q_i \int dx^0 A^{(i)}_0 \, ,
 \ee
 to the constant mode of $A^{(i)}_0$. On the other hand the magnetic
 charge $p_i$ is defined in terms of the 
asymptotic form of $F^{(i)}_{\theta\phi}$:
 \be \label{eapa2}
 F^{(i)}_{\theta\phi}={p_i\over 4\pi}\, \sin\theta\, .
 \ee
 We begin by considering an elementary 
string wrapped once along $S^1$. 
In the presence of such a string there is a coupling
 \be \label{eapa3}
 {1\over 4\pi\alpha'} \int d\xi^0 \int d\xi^1 \, 
 \varepsilon^{\alpha\beta}\, B_{MN}\, \p_\alpha X^M \p_\beta X^N 
 \, ,
 \ee
where $\xi^0$ and $\xi^1$ are the world-sheet coordinates and
$\varepsilon^{\alpha\beta}$ is the totally anti-symmetric tensor on
the world-sheet with $\varepsilon^{10}=1$. 
In the static gauge $X^0=\xi^0$, $X^9=\xi^1$,
this gives a coupling:
\be \label{eapa4}
{1\over 2\pi\alpha'} B_{90} \int d x^0 \int dx^9 =  {1\over 4} \int dx^0
B_{90}\, ,
\ee
where we have used $\alpha'=16$ and the fact that $x^9$ has periodicity 
$2\pi\sqrt{\alpha'}=8\pi$. 
Using the identification $B_{9\mu}=2\, A^{(3)}_\mu$ given in 
\refb{e6}
we can rewrite \refb{eapa4} as
\be \label{eapa5}
{1\over 2} \int dx^0\, A^{(3)}_0\, .
\ee
Comparing with \refb{eapa1} we see that an 
elementary string wound once 
along $S^1$ carry half a unit of $q_3$ charge. Thus 
for an elementary string wound $w$ times along $S^1$, we have
\be \label{eapa6}
q_3={1\over 2} \, w\, .
\ee
Since T-duality along the circle $S^1$ interchanges the momentum and 
winding along $S^1$ and also the gauge fields $A^{(1)}$ and $A^{(3)}$, we get
\be \label{eapa7}
q_1 = {1\over 2}\, n\, .
\ee
Finally, symmetry under the exchange $S^1\leftrightarrow \wt S^1$ gives
\be \label{enx1}
q_2={1\over 2} \, \wt n, \qquad q_4 = {1\over 2}\, \wt w\, .
\ee

We now turn to the normalization of the magnetic charges. Asymptotically
a Kaluza-Klein 
monopole solution associated with the circle $\wt S^1$ corresponds to a 
(9+1) dimensional background:
\be \label{ekk1}
ds_{10}^2 = -dt^2 + a \, (dx^8 + 2(1-\cos\theta)d\phi)^2 + dr^2+ r^2 
(d\theta^2+\sin^2\theta d\phi^2) + b\, (dx^9)^2 + d\vec y^2\, ,
\ee
where $d\vec y^2$ denotes the metric on $\MM$ and $a$, $b$ are constants. 
The coefficient in front of 
$(1-\cos\theta)$ has been fixed so that at $\theta=\pi$, the 
$\phi\to\phi+2\pi$ transformation can be compensated by translating $x^8$ 
by its periodicity $8\pi$. Otherwise the $\phi$ circle will not shrink to 
a point at $\theta=\pi$. Using \refb{e6} we see that for large $r$ this 
corresponds to
\be \label{ekk2}
A^{(2)}_\phi = (1-\cos\theta)\, ,
\ee
and hence
\be \label{ekk3}
F^{(2)}_{\theta\phi} = \sin\theta\, .
\ee
{}From \refb{e0} it follows that the above configuration has 
$p_2=4\pi$. Thus 
$\wt N$ Kaluza-Klein monopoles associated with the circle $\wt S^1$ will 
have
\be \label{ekk4}
p_2=4\pi\, \wt N\, .
\ee
Since T-duality associated with $\wt S^1$ exchanges $\wt N$ and $\wt W$ 
and also $A^{(2)}_\mu$ and $A^{(4)}_\mu$, it follows that
\be \label{ekk5}
p_4=4\pi\, \wt W\, .
\ee
Finally, using the $S^1\leftrightarrow \wt S^1$ symmetry we get
\be \label{enx2}
p_1=4\pi\, N, \qquad p_3=4\pi\, W\, .
\ee

\end{document}